\documentclass[submission, PhysCodeb]{SciPost_better_arXiv}
\hypersetup{
    colorlinks,
    linkcolor={red!50!black},
    citecolor={blue!50!black},
    urlcolor={blue!80!black}
}

\usepackage[bitstream-charter]{mathdesign}
\usepackage{listings}
\definecolor{comment}{rgb}{0,0.3,0}
\definecolor{identifier}{rgb}{0.0,0,0.3}

\usepackage{booktabs}
\DeclareSymbolFont{usualmathcal}{OMS}{cmsy}{m}{n}
\DeclareSymbolFontAlphabet{\mathcal}{usualmathcal}

\newcommand{\hoppet}{{\sc hoppet}}
\newcommand{\disent}{{\tt disent}}
\newcommand{\disorder}{{\tt disorder}}
\newcommand{\provbfh}{{\tt proVBFH}}
\newcommand{\fastjet}{{\tt fastjet}}
\newcommand{\lhapdf}{{\tt LHAPDF}}
\newcommand{\nnlojet}{NNLO{\sc{jet}}}
\newcommand{\disaster}{{\tt disaster++}}
\newcommand{\nlojet}{{\tt nlojet++}}
\newcommand{\as}{\alpha_{\mathrm{s}}}
\newcommand{\dd}{\mathrm{d}}
\newcommand{\NC}{\mathrm{NC}}
\newcommand{\CC}{\mathrm{CC}}
\newcommand{\GEV}{\,\mathrm{GeV}}
\newcommand{\NNNLO}{N$^3$LO}
\newcommand{\ttt}[1]{\texttt{#1}}

\newcommand{\repolink}[2]{\href{https://github.com/alexanderkarlberg/disorder/blob/master/#1}{\ttt{#2}}}
\newcommand{\masterlink}[1]{\repolink{#1}{#1}}
\newcommand{\email}[1]{\href{mailto:#1}{#1}}

\begin{document}
\begin{flushright}
CERN-TH-2023-229
\end{flushright}
\begin{center}{\Large \textbf{
disorder: Deep inelastic scattering at high orders\\
}}\end{center}

\begin{center}
Alexander Karlberg
\end{center}

\begin{center}
CERN, Theoretical Physics Department, CH-1211 Geneva 23, Switzerland
\\[0.5cm]
{\small \sf \email{alexander.karlberg@cern.ch}}
\end{center}



\section*{Abstract}
{\bf
We present a Fortran 77/95 code capable of computing QCD corrections
in deep inelastic scattering (DIS). The code uses the
Projection-to-Born method to augment an existing $\mathcal{O}(\as^2)$
dijet DIS code, thereby obtaining predictions for photon-mediated
neutral-current single-jet DIS production in the laboratory frame. The
code is lightweight and fast, and yet includes the most common
functionalities found in typical perturbative QCD programs, like
automatic renormalisation and factorisation scale uncertainties,
options to run and combine multiple seeds, and interfaces to
\fastjet{} and \lhapdf{}. Due to the underlying \disent{} and
\hoppet{} codes, the program also provides stable results in the
infrared, relevant for extracting logarithmic coefficients for
analytic resummations, and access to the massless DIS structure
functions and (reduced) cross sections up to $\mathcal{O}(\as^3)$.}

\vspace{10pt}
\begin{center}\url{https://github.com/alexanderkarlberg/disorder}\end{center}
\vspace{10pt}
\noindent\rule{\textwidth}{1pt}
\tableofcontents\thispagestyle{fancy}
\noindent\rule{\textwidth}{1pt}
\newpage
\section{Introduction}
\label{sec:intro}
Deep inelastic scattering (DIS) is arguably one of the best understood
processes in perturbative QCD. It is one of few processes for which
there exists an exact factorisation
theorem~\cite{Collins:1987pm,Collins:1989gx}, and the massless
unpolarised hard perturbative coefficients have been computed through
an impressive three
loops~\cite{SanchezGuillen:1990iq,vanNeerven:1991nn,Zijlstra:1992qd,Zijlstra:1992kj,vanNeerven:1999ca,vanNeerven:2000uj,Moch:1999eb,Moch:2004xu,Vermaseren:2005qc,Vogt:2006bt,Moch:2007rq,Davies:2016ruz,Blumlein:2022gpp}
with progress on the four loop results presented recently in
Ref.~\cite{Moch:2022frw}. Together with the three-loop results for the
DGLAP splitting functions~\cite{Moch:2004pa,Vogt:2004mw,Blumlein:2021enk}
and the four-loop
$\beta$-function~\cite{vanRitbergen:1997va,Czakon:2004bu} for the
running coupling this allows for the determination of the massless
proton structure functions at the next-to-next-to-next-to-leading
order (\NNNLO{}).\footnote{\label{n3lo-note}Technically the four-loop
splitting functions are needed to claim this accuracy. Recent progress
in determining those and the three-loop massive operator matrix
elements, needed for the variable-flavour-number
scheme~\cite{Buza:1996wv}, can be found in
Refs.~\cite{Moch:2021qrk,Falcioni:2023luc,Falcioni:2023vqq,Gehrmann:2023cqm,Falcioni:2023tzp,Moch:2023tdj,Gehrmann:2023iah,Falcioni:2024xyt}
and
\cite{Bierenbaum:2009mv,Kawamura:2012cr,ABLINGER2014263,Ablinger:2014vwa,Ablinger:2014nga,Ablinger:2022wbb,Ablinger:2023ahe,Ablinger:2024xtt}
respectively. Phenomenologically the first few moments which have been
computed therein, with some exact pieces also being known, are however
in all likelihood enough to claim \NNNLO{}. } The structure functions
can be combined with an exclusive next-to-next-to-leading order (NNLO)
dijet DIS computation to obtain fully differential \NNNLO{}
predictions for single-jet DIS production, as was done by the
\nnlojet{}
collaboration~\cite{Currie:2016ytq,Currie:2018fgr,Gehrmann:2018odt}. This
computation uses the Projection-to-Born (P2B) method, which was first
introduced in the context of NNLO Vector Boson Fusion (VBF)
production~\cite{Cacciari:2015jma}, and has since also been extended
to \NNNLO{} colour-singlet production in proton-proton
collisions~\cite{Chen:2021isd,Chen:2021vtu}.

Despite this impressive theoretical progress, there are few publically
available computer codes from which one can obtain fast and reliable
high-order differential cross section predictions for DIS. This paper
and associated Fortran code seeks to address that.

One of the advantages of the P2B method is that it is rather agnostic
towards the details of the underlying exclusive computation, and it is
hence possible to augment the validity of existing fixed-order
codes. Historically the most successful fixed-order codes have been
{\tt MEPJET}~\cite{Mirkes:1995ks}, \disent{}~\cite{Catani:1996vz},
\disaster{}~\cite{Graudenz:1997gv}, and \nlojet{}~\cite{Nagy:2001xb}
which are all next-to-leading order (NLO) accurate for DIS dijet
production. Discrepancies between \disent{} and fixed-order
coefficients from analytical resummation were initially observed in
Refs.\cite{Antonelli:1999kx,Dasgupta:2002dc}, and were only recently
understood to be due to a bug in one of the dipole terms for the gluon
channel in \disent{}~\cite{Borsa:2020ulb,Borsa:2020yxh}.

With the bug fixed we can use \disent{} as the underlying exclusive
dijet NLO code together with the NNLO structure functions
from~\hoppet{}~\cite{Salam:2008qg,BertoneKarlberg},\footnote{The
combination of \disent{} and \hoppet{} naturally leads to a Fortran
code. For a C++ alternative, based on publically available codes, one
could have started from either \disaster{} or \nlojet{} and used the
structure functions as implemented in
APFEL++~\cite{Bertone:2013vaa,Bertone:2017gds}} to obtain fully
differential single-jet DIS predictions. Advantages of using \disent{}
are its well-known efficiency and numerical stability. The resulting
program is dubbed \disorder{} keeping in the spirit of previous names
for fixed-order DIS codes.\footnote{{\tt
  dispatch}~\cite{Dasgupta:2002dc} deserves an honorable mentioning in
this context.} Since the program relies on \hoppet{} for the structure
functions, the inclusive DIS cross section can actually be obtained
one order higher, as the structure functions were implemented at
\NNNLO{} already in the context of the \provbfh{}
codes~\cite{Cacciari:2015jma,Dreyer:2016oyx,Dreyer:2018qbw,Dreyer:2018rfu}
and are publically available since {\tt v1.3.0} of
\hoppet{}~\cite{hoppetv130}.\footnote{\hoppet{} implements all known
approximate and exact ingredients at this order - see footnote
\ref{n3lo-note}.} The \disorder{} program therefore has two more or
less separate use cases: 1. The computation of fully differential
$\mathcal{O}(\as^2)$ photon-mediated neutral-current (NC) massless DIS
and 2. The computation of inclusive and charged-current (CC) massless
DIS at \NNNLO{} accuracy.

DIS is peculiar in-so-far that despite having a final-state parton
already at the lowest order, the existence of a non-trivial jet, at
this order, is highly frame-dependent. The majority of jet analyses
are performed in the Breit-frame, as was for instance the case at
HERA~\cite{ZEUS:2006xvn,H1:2009pqp,ZEUS:2010vyw,H1:2014cbm,Baghdasaryan:2015yha,H1:2016goa,H1:2024pvu,H1:2024aze}. In
this frame the incoming proton collides head-on with the photon, and
the resulting jet has zero transverse momentum. At NLO, the
real-emission diagram can instead give rise to two jets with equal and
opposite transverse momentum, provided the two partons are not
clustered together. As a consequence, the first non-trivial jet
process that can be described in the Breit-frame is dijet
production. In contrast, in the laboratory frame (or virtually any
other frame than the Breit-frame) there will always be at least one
jet present.\footnote{Although the Breit-frame is used in most
analyses, there exist many laboratory frame measurements as
well~\cite{E665:1992xqj,E665:1993vlk,H1:1993rdo,ZEUS:1995tgg,ZEUS:1995wzd,ZEUS:2005ukc}
including the very recent ZEUS measurement~\cite{ZEUS:2024mhu}.}  The
P2B method therefore only meaningfully augments the \disent{} program
in the laboratory frame, where single-jet production is
well-defined. Hence, when analysing the output of \disorder{} in the
Breit-frame it will essentially be identical to that which one would
obtain by running a stand-alone version of \disent{}, with the
addition that the inclusive NNLO cross section is computed correctly
at the same time.

The \disorder{} program is designed to be user-friendly with a simple
command line interface. It prints cross sections and all important
run-parameters to the screen and disk, allowing a user to acquire
cross sections with very little effort. It comes with an interface to
\fastjet{}~\cite{Cacciari:2011ma} and \lhapdf{}~\cite{Buckley:2014ana}
and uses the histogramming package from the {\tt
  POWHEG-BOX}~\cite{Alioli:2010xd} for easy analysis. The code can
also compute renormalisation and factorisation scale uncertainties
on-the-fly. For inclusive cross sections the code provides results in
a matter of seconds\footnote{The majority of this time is taken up by
the structure function initialisation inside \hoppet{} rather than the
integration of the cross section.} even at \NNNLO{} with integration
uncertainties typically below the permille level. For exclusive
quantities the code can be run on a laptop at NLO and depending on the
analysis, and the laptop, even at NNLO.

It should be pointed out that the code provides little theoretical
advance on its own. As described above, the structure functions
themselves have been known for a while, and the P2B method has already
been applied to DIS, even one order higher than here.\footnote{It has
also been applied in polarised DIS in
Refs.~\cite{Borsa:2020ulb,Borsa:2020yxh,Borsa:2021afb,Borsa:2022irn,Borsa:2022cap}.}
In addition only photon-mediated DIS can be computed fully
differentially at the moment. It is however the author's opinion that
with the renewed interest in DIS due to the upcoming
Electron-Ion-Collider (EIC)~\cite{AbdulKhalek:2022hcn},
well-maintained public code is extremely valuable for both the
experimental and theoretical communities, and that providing
documentation in the form of this article will enable the wide use of
the code.

The paper is structured as follows: In section~\ref{sec:dis} we review
the DIS process and kinematics as implemented in \disorder{}. In
section~\ref{sec:running} we provide details on how to run \disorder{}
and in section~\ref{sec:results} we show a few results from the
program. Finally we conclude in section~\ref{sec:conclusion}.

\section{Basics of the DIS process}
\label{sec:dis}
In this section we first give some standard definitions for kinematics
and cross sections in DIS. Along the way we specify the conventions
that are used in the \disorder{} code, and provide some details on the
P2B method as applied there.

At leading order (LO) the DIS process is
the scattering of a massless (anti-)quark $q$ off a massless
\mbox{(anti-)lepton} $l$ via the exchange of a photon or electroweak
gauge boson $V$ of virtuality $Q^2$. Denoting the external
four-momenta by $k_i$ (incoming lepton), $k_f$ (outgoing lepton),
$p_i$ (incoming quark), and $p_f$ (outgoing quark) we can define the
Lorentz invariant DIS variables $x$, $Q^2$, and $y$, given by
\begin{align}
  Q^2 = -q^2 = -(k_i - k_f)^2, \qquad  x = \frac{Q^2}{2 P \cdot q},
  \qquad  y = \frac{P \cdot q}{P \cdot k_i} = \frac{p_i \cdot q}{p_i \cdot k_i} ,
  \label{eq:dis-variables}
\end{align}
where $P$ is the hadron four-momentum. As can be seen these kinematics
are fully specified by the hadron and lepton momenta. This is also
true beyond LO.

DIS is most often analysed in the Breit-frame which is specified by
requiring that $2 x \vec{P} + \vec{q} = 0$.\footnote{For the explicit
transformation between lab- and Breit-frames we follow Appendix 7.11
in Ref.~\cite{Devenish:2004pb}.} In this frame the mediated vector
boson has zero energy component and is anti-aligned with the incoming
parton. Explicitly in \disorder{} the Breit-frame at LO is
\begin{align}
  \label{eq:BreitLO}
k_i &= \frac{Q}{2}\left(\frac{2-y}{y},\frac{2\sqrt{1-y}}{y},0,-1 \right)\,, \quad  p_i = \frac{Q}{2}\left(1,0,0,+1 \right)\,, \\ \notag
k_f &= \frac{Q}{2}\left(\frac{2-y}{y},\frac{2\sqrt{1-y}}{y},0,+1 \right)\,, \quad  p_f = \frac{Q}{2}\left(1,0,0,-1 \right)\,, 
\end{align}
where four-momenta are given as $(E,p_x,p_y,p_z)$. The resulting
vector $q$ is hence given by $(0,0,0,-Q)$. It is clear from
eq.~\eqref{eq:BreitLO} that the outgoing parton in the Breit-frame has
zero transverse momentum. In the laboratory frame we align the parton
with the positive $z$-axis and hence the lepton with the negative
\begin{align}
\tilde{k}_i &= E_\mathrm{l}\left(1,0,0,-1 \right), \quad  \tilde{p}_i = x E_\mathrm{h}\left(1,0,0,+1 \right)\,, 
\end{align}
where $E_\mathrm{h}$ is the energy of the incoming hadron. Using the
definitions of eq.~\eqref{eq:dis-variables} one finds that in the lab
frame
\begin{equation}
  \tilde{q} = \left(y(E_\mathrm{l} - x E_\mathrm{h}), -Q\sqrt{1-y},0, -y(E_\mathrm{l} - x E_\mathrm{h})\right)\,, 
\end{equation}
and the outgoing momenta then simply follow from momentum conservation
\begin{equation}
  \label{eq:LabLO}
  \tilde{k}_f = \tilde{k}_i - \tilde{q}, \quad   \tilde{p}_f = \tilde{p}_i + \tilde{q}\,. 
\end{equation}
Conversely, in the laboratory frame, the outgoing parton always has
transverse momentum of magnitude $Q\sqrt{1-y}$. In \disorder{} both
the Breit-frame and laboratory frame momenta can be accessed by the
analysis at the same time.

The inclusive cross section for DIS can be split into a NC
contribution, from $e^\pm p \to e^\pm + X$ scattering, and a CC
contribution from $e^\pm p \to \nu + X$ scattering.\footnote{Here
$\nu=\{\nu_e,\bar{\nu}_e\}$ for $e^-$ and $e^+$ respectively. One can
of course also consider incoming neutrinos which does not change the
discussion here, except adding a factor $2$ in eq.~\eqref{eq:CCsigma}
and changing the vector and axial-vector couplings to $v_e = \frac12$
and $a_e=-\frac12$.} The unpolarised NC cross section can
be written as
\begin{equation}
\frac{\dd\sigma_{\NC}^\pm}{\dd x \dd Q^2} =   \frac{2\pi\alpha^2}{xQ^4} \left[y_+ F_2^{\NC} \mp y_- x F_3^\NC - y^2 F_L^\NC\right],\,
\label{eq:NCsigma}
\end{equation}
where $y_\pm=1\pm(1-y)^2$, $\alpha$ is the fine structure constant and
$F_i^\NC$ can be expressed in terms of the usual proton structure
functions
\begin{align}
  F_{i}^\NC &= F_{i}^\gamma  - v_e \Gamma_{\gamma Z} F_{i}^{\gamma Z} + (v_e^2 + a_e^2)\Gamma_{Z} F_{i}^{Z},\, \quad i =2,L \\ \notag
  F_{3}^\NC &= - a_e \Gamma_{\gamma Z} F_{3}^{\gamma Z} + 2v_e a_e\Gamma_{Z} F_{3}^{Z}\,.
\end{align}
Here $v_e = -\frac12 + 2 \sin^2\theta_W$ and $a_e=\frac12$ are the
vector and axial-vector couplings respectively, $M_Z$ is the Z boson
mass, $\Gamma_{\gamma Z} = \frac{Q^2}{\sin^2 2\theta_W(Q^2+M_Z^2)}$,
and $\Gamma_Z=\Gamma_{\gamma Z}^2$. $\theta_W$ is the weak mixing
angle. In \disorder{} the electroweak parameters are fixed by
$\alpha$, $M_W$, and $M_Z$ through the tree-level relations
\begin{align}
  \sin^2\theta_W = 1-\frac{M_W^2}{M_Z^2}, \qquad G_F = \frac{\pi\alpha}{\sqrt{2}M_W^2\sin^2\theta_W},
\end{align}
with $G_F$ the Fermi constant. Similarly we define the unpolarised CC
cross section as
\begin{equation}
\frac{\dd\sigma_{\CC}^\pm}{\dd x \dd Q^2} =   \frac{2\pi\alpha^2}{xQ^4}\left[\frac{Q^2}{4\sin^2\theta_W(M_W^2 + Q^2)}\right]^2 \left[y_+ F_2^{\CC} \mp y_- x F_3^\CC - y^2 F_L^\CC\right],\,
\label{eq:CCsigma}
\end{equation}
where $M_W$ is the mass of $W$ boson and the CC structure functions
are now simply given by the $W$ ones
\begin{equation}
  F_2^\CC = F_2^{W^\pm}, \quad   F_L^\CC = F_L^{W^\pm}, \quad   F_3^\CC = F_3^{W^\pm}\,.
\end{equation}
The exact definitions of all proton structure functions inside
\hoppet{} up to \NNNLO{} can be found in
Refs.~\cite{Salam:2008qg,hoppetv130,BertoneKarlberg}. The tabulation of the
structure functions inside \hoppet{}, as used in \disorder{}, has a
relative numerical precision of $10^{-4}$ for most values of $x$,
which in turn limits the accuracy which can be obtained to the same
order.

In DIS it is also customary to define the dimensionless
\emph{reduced} NC and CC cross sections by~\cite{H1:2012qti}
\begin{align}
  \label{eq:reducedsigma}
  \tilde{\sigma}_\NC^\pm(x,Q^2) &= \frac{xQ^4}{2\pi\alpha^2}\frac{1}{y_+}\frac{\dd\sigma_{\NC}^\pm}{\dd x \dd Q^2}\,, \notag \\
  \tilde{\sigma}_\CC^\pm(x,Q^2) &= \frac{8\sin^4\theta_W x}{\pi\alpha^2}\left[M_W^2+Q^2\right]^2\frac{\dd\sigma_{\CC}^\pm}{\dd x \dd Q^2}\,.
\end{align}
\disorder{} provides direct access to all the cross sections in
eqs.~\eqref{eq:NCsigma},~\eqref{eq:CCsigma},
and~\eqref{eq:reducedsigma} at \NNNLO{} accuracy. In principle one can
also access the NC and CC structure functions, although they are
currently only computed as an intermediate step to construct the cross
sections and are not accessible in the user analysis.

\subsection{A note on the phase space}
\label{sec:phasespace}
Since the Born phase space is fully constrained by specifying any two
of the three DIS variables $x$, $y$, and $Q^2$ there exists more than
one double-differential cross section. The convention in \disorder{}
is to always return $\frac{\dd\sigma^2}{\dd x \dd Q^2}$ (in
pb/$\GEV^2$) regardless of which of the two variables are fixed. The
user can easily convert to $\frac{\dd\sigma^2}{\dd x \dd y}$ or
$\frac{\dd\sigma^2}{\dd y \dd Q^2}$ by supplying a factor
$\frac{Q^2}{y}$ or $\frac{x}{y}$ respectively. If only one of $x$,
$y$, or $Q^2$ is specified then the appropriate single differential
cross section is returned, i.e.~$\frac{\dd\sigma}{\dd x}$ in pb,
$\frac{\dd\sigma}{\dd y}$ in pb, or $\frac{\dd\sigma}{\dd Q^2}$ in
pb/$\GEV^2$, integrated over the other two variables. If none of the
variables are fixed then the total cross section integrated over $x$,
$y$, and $Q^2$ is returned in pb.

\subsection{Applying P2B}
\label{sec:P2B}
The structure functions are by definition inclusive in all radiation
and can therefore only provide predictions for quantities, like the
inclusive cross sections, which depend on the Born kinematics of
eq.~\eqref{eq:dis-variables} only. If we instead evaluate the
structure functions on an observable sensitive to emissions, e.g. the
transverse momentum of the hardest jet in the laboratory frame, we see
that this will not give the right answer, as the real emissions are
not included with their correct-kinematics. In fact, when computing
the coefficient functions that enter the structure functions, the
real-emission diagrams are explicitly projected onto the Born
kinematics.

The P2B method lifts this restriction by effectively replacing the
Born-kinematics real-emission contributions in the structure functions
with the correct kinematics ones. In practice whenever \disent{}
returns an event with some weight, we bin it once according to the
true kinematics, and again projecting the kinematics to the underlying
Born changing the sign of the weight. This last term, upon
integration, will exactly cancel the real contribution in the
structure functions, whereas the first term will provide the correct
real matrix element. A detailed discussion of the method can be found
in section 2 of Ref.~\cite{Currie:2018fgr}.

As mentioned already in the Introduction, this procedure is trivial
when applied to the Breit-frame kinematics. In this frame the outgoing
parton has zero transverse momentum, and will therefore not contribute
to any jet-sensitive observable, like the well-known thrust
event-shape~\cite{Antonelli:1999kx}. In the laboratory frame however,
the jet kinematics are non-trivial even at Born-level,
cf. eq.~\eqref{eq:LabLO}, and the P2B method as applied here will
correctly describe single-jet production in this frame. The
projections themselves are trivial because, as outlined in the section
above, the Born kinematics are fully specified at all orders by the
lepton (and proton) momenta, as given in
eqs.~\eqref{eq:BreitLO}--\eqref{eq:LabLO}. Hence computing the
projections adds very little computational effort to the cross section
calculation.

At this point it is worth reminding the reader that \disent{} only
includes the photon-mediated NC, and hence \disorder{} only provides
exclusive predictions for this channel. It should be possible to
extend the code to include Z-mediation (including interferences) and
CC, but we leave this for future work.

In principle P2B could be applied to the \disent{} code without any
major modifications besides the bug fix mentioned in the introduction,
which has been implemented here. However a few significant
modifications were introduced to allow for a more flexible integration
of \disent{} into \disorder{}. The version of \disent{} that we
include in \disorder{} is based on the version which can be found in
{\tt dispatch}~\cite{Dasgupta:2002dc}, and which already included some
minor modifications. In addition to the modifications present there,
we have also introduced $\alpha$ as an input parameter, whereas before
it was fixed to $1/137$. A number of parameters can also now be set on
the command line as described below. Besides coding the actual
interface to \disent{}, the biggest modification introduced in
\disorder{} is that the {\tt KPFUNS} subroutine now returns an array
of weights corresponding to varying the factorisation scale by a
factor two up and down. This allows for much faster evaluation of
scale uncertainties compared to running the program three separate
times. Additionally \disent{} can now also use any scale defined in
terms of $x$, $y$, and $Q$, and not just some multiple of $Q$. We
provide a few different scale choices as documented below, but more
can easily be implemented.

\section{Running \disorder{}}
\label{sec:running}
In this section we give instructions on how to compile and run
\disorder{}, giving a few examples of the use of the most common
command line arguments. The code itself can be obtained from

\begin{center}\url{https://github.com/alexanderkarlberg/disorder}\end{center}

\subsection{Compiling and prerequisites}
A user should start by inspecting \masterlink{README.md}. To compile
\disorder{} both \hoppet{} ({\tt v1.3.0} or later) and \lhapdf{}
(tested with {\tt v6.5.4})~\cite{Buckley:2014ana} have to be installed
on the machine. If both are installed in a location in the {\tt
  \$PATH} it is enough to run
\begin{lstlisting}
  mkdir build && cd build
  cmake ..

  make [-j]
\end{lstlisting}
from the main directory. This will create an executable \disorder{}
along with two auxiliary executables, {\tt mergedata} and {\tt
  getpdfuncert}. For non-standard installation of \hoppet{} and
\lhapdf{} the paths can be specified like this
\begin{lstlisting}
  cmake -DHOPPET_CONFIG=/path/to/hoppet-config -DLHAPDF_CONFIG=/path/to/lhapdf-config
\end{lstlisting}
where the path should include the config-file itself (i.e. {\tt
  /usr/local/bin/hoppet-config}). By default \fastjet{} is not linked
and only a skeleton analysis
(\masterlink{analysis/simple\_analysis.f}) is compiled. To link
\fastjet{} run
\begin{lstlisting}
  cmake -DNEEDS_FASTJET=ON [-DFASTJET_CONFIG=/path/to/fastjet-config]
\end{lstlisting}
where the path to {\tt fastjet-config} only needs to be specified if
it is not in the user's {\tt \$PATH}. To compile a different analysis
the user should first put it in the \masterlink{analysis} directory (here we
assume it to be called {\tt my\_analysis.f}), and then pass it to
{\tt cmake} through
\begin{lstlisting}
  cmake -DANALYSIS=my_analysis.f
\end{lstlisting}
The program has been found to compile on a Linux machine using the
{\tt gfortran v11.4.0} compiler and also on various MacOS systems.

\subsubsection{The analysis framework}
The code uses the {\tt POWHEG-BOX} analysis framework, with some minor
modifications. A few example analyses are included in the \masterlink{analysis} directory. Any new analysis should be put here and the
name of the analysis should be passed to {\tt cmake} as described
above. There are two mandatory routines in the analysis file, {\tt
  define\_histograms} and {\tt user\_analysis}. In the first routine
one should define histograms like this (there are also routines
available to book histograms with varying bins sizes)
\begin{lstlisting}
  call bookupeqbins('string_name', binsize, min, max)
\end{lstlisting}
The {\tt user\_analysis} routine takes as input
\begin{lstlisting}
  integer n
  double precision dsig(maxscales), x, y, Q2
\end{lstlisting}
where {\tt n} is the number of initial plus final state particles,
{\tt dsig} is the weights computed by \disorder{} and {\tt maxscales}
is the maximum number of scales which is supported (currently 7). {\tt
  x}, {\tt y}, and {\tt Q2} are the DIS variables.

Through the module {\tt mod\_analysis} the analysis has access to two
arrays of momenta {\tt pbreit(0:3,1:n)} and {\tt plab(0:3,1:n)} in
which the Breit and laboratory frame momenta are stored
respectively. The first entry is the incoming lepton, the second the
incoming parton, the third the outgoing lepton and the rest outgoing
partons. The output of the analysis will be saved to the disk as
outlined below.  In the {\tt user\_analysis} routine the user should
perform their analysis and fill histograms like this
\begin{lstlisting}
  call filld('string_name', obs_value, dsig)
\end{lstlisting}
where {\tt obs\_value} is the value of the observable to be binned, and
{\tt dsig} is the associated array of weights. 

In the {\tt aux} folder one may also find a
script called {\tt mergedata} that can perform various manipulations
of the datafiles. In particular
\begin{lstlisting}
  ./mergedata 1 {list of statistically equivalent files}
\end{lstlisting}
will take the average of all files and produce the file {\tt fort.12}
with the result. Running the script without any arguments will result
in a list of possible uses of the script. The {\tt mergedata} script
is taken from the {\tt POWHEG-BOX} as well.

\subsection{Inclusive mode}
\label{sec:Inclusive}
The syntax for running the program is
\begin{lstlisting}
  ./disorder -pdf LHAPDF_name [options]
\end{lstlisting}
Running the program without any other options than {\tt -pdf} will
compute the total inclusive cross section above the minimum $Q$ value
accessible in the PDF, and using default parameters everywhere. The
user can get a list of most parameters which can be specified on the
command line by running
\begin{lstlisting}
  ./disorder -help
\end{lstlisting}
Here we describe the most common flags, but for a complete list of all
parameters, and their use, the user should look through the file
\masterlink{src/mod\_parameters.f90}.

The program allows the user to specify limits on $x$, $y$, and $Q$
through {\tt -xmin}, {\tt -xmax}, {\tt -ymin}, {\tt -ymax}, {\tt
  -Qmin}, {\tt -Qmax}, or to fix them through the options {\tt -x},
{\tt -y}, {\tt -Q}. For instance to compute the cross section at $Q=20
\GEV$ and $x>0.01$ one would run
\begin{lstlisting}
  ./disorder -Q 20 -xmin 0.01 -pdf LHAPDF_name
\end{lstlisting}
The program will perform a Monte Carlo integration in the ranges
specified, using the integrator VEGAS~\cite{Lepage:1977sw}. If the
phase space is fully constrained by fixing two of either $x$, $y$, and
$Q$ the program simply evaluates one point and returns the answer.

One can further specify the energy of the incoming lepton through {\tt
  -Elep}, the incoming hadron through {\tt -Ehad}. By default the
lepton is taken to be an electron but specifying {\tt -positron} on
the command line will change that. To use an incoming neutrino the
user should specify {\tt -neutrino}. If {\tt -positron} is also
specified the incoming lepton will be an anti-neutrino. The code
computes the photon-mediated NC cross section only by default. To
include the $Z$ one can specify {\tt -includeZ} and to include CC
processes one can specify {\tt -CC}. The inclusion of NC processes can
also be controlled through the {\tt -NC} flag. In fact, all logical
flags can be prefixed by ``{\tt{no}}'' to turn them off. Hence the
below command line would run the program with CC processes only using
a positron
\begin{lstlisting}
  ./disorder -Q 20 -xmin 0.01 -noNC -CC -positron -pdf LHAPDF_name
\end{lstlisting}

The order of the calculation is by default NNLO but can be specified
with one of the flags {\tt -lo/-nlo/-nnlo/-n3lo}. If the user wants to
compute PDF uncertainties, the flag {\tt -pdfuncert} should be
given. This flag will make \disorder{} loop over all the members in
the PDF, and combine their errors according to the routine {\tt
  getpdfuncertainty}~\cite{Watt:2011kp}, native to LHAPDF. This also
means that the program is slowed down proportionally to the number of
PDF members (although the VEGAS grid only gets computed once and then
stored so that the runs are fully correlated). If the PDF also
includes $\as$ variations these are included in the PDF uncertainty by
default. If the user wants the PDF and $\as$ uncertainties
independently, then the flag {\tt -alphasuncert} should be
specified. Some care should be taken here, as in practice the code
assumes that the $\as$ variations are contained in the last two PDF
members, and simply separates them from the rest.

Renormalisation, $\mu_R$, and factorisation, $\mu_F$, scale
uncertainties can be included by specifying the flag {\tt
  -scaleuncert}. \disorder{} uses the vector boson virtuality, $Q^2$,
as its default central scale, but can use any scale as long as it is
defined in terms of $x$, $y$, and $Q$. A few central scales, $\mu$,
are currently implemented and can be accessed through the {\tt
  -scale-choice} flag. The options are {\tt 0}: $M_Z$, {\tt 1}
(default): $Q^2$, {\tt 2}: $Q^2(1-y)$, {\tt 3}:
$\frac{Q^2(1-x)}{x}$. The program will then compute a standard 7-point
scale variation varying this scale by a factor of two up and down
keeping $1/2\le \mu_R/\mu_F\le 2$. On the screen the envelope of all 7
runs will be printed. The user can in principle also carry out
arbitrary variations in individual runs by specifying {\tt -xmur} and
{\tt -xmuf} on the command line. Here {\tt xmur} is the ratio of
$\mu_R/\mu$ and similarly for {\tt xmuf}.\footnote{In practice it is
faster to use the {\tt -scaleuncert} flag as the program will only
recompute what is needed for the variations rather than do a full
event. However, at \NNNLO{}, in general, this is not true, due to the
fact that the number of tables needed in \hoppet{} to carry out
on-the-fly scale variations increases dramatically at this order,
compared to using a fixed ratio of $Q$. Given how fast the code is, it
is still the author's opinion that it is more convenient to use the
on-the-fly variations.}

Finally the random seed can be set with the {\tt iseed} flag. When the
program terminates it will print results to screen but also save a
number of files depending on the exact input. There is always a file
called {\tt xsct\_nnlo\_seed0001.dat}, where the {\tt nnlo} and {\tt
  seed0001} parts will vary depending on the order and seed, which
contains a summary of the run, including the total and reduced cross
sections. 

The output of the analysis is printed to a number of {\tt .dat} files,
the number depending on the input, prefixed by {\tt
  disorder\_nnlo\_seed0001\_pdfmem000} where again the exact prefix
will depend on the input. The output name also contains information on
the seed, the PDF member and the renormalisation and factorisation
scales if {\tt -scaleuncert} is on. The user can specify a prefix to
be added to all the files through the {\tt -prefix} flag.

Since it is often not necessary to run an analysis in the inclusive
mode, the user can simply turn this off with the {\tt -no-analysis}
flag.

Finally the user can control the number of VEGAS integration points
through the two flags {\tt -ncall1} and {\tt -ncall2}. The first flags
controls the number of points to use to set up the grids. Since the
phase space is not complicated it is rarely necessary to increase the
default of 10000. The second flag controls the number of points that
are used for the actual integration. If one is not running an
analysis, the default number of 100000 should give results that have
better than permille level accuracy. If the user wishes for instance
to bin the cross section in fine bins of $x$ and $Q^2$, this number
will most likely have to be increased.

\subsection{P2B mode}
\label{sec:P2Bmode}
To turn on P2B it is enough to specify {\tt -p2b} on the command
line. Many of the flags described above can also be specified in this
mode, with a few limitations
\begin{itemize}
\item {\tt -CC}, {\tt -neutrino}, and {\tt -includeZ} are not supported
\item {\tt -n3lo} is not supported
\item {\tt -pdfuncert} and {\tt -alphasuncert} are not supported
\end{itemize}
Importantly, it \emph{is} possible to run with the {\tt -scaleuncert}
flag which leads to a significant reduction in run-time compared to
doing seven separate runs. When running in P2B mode it is furthermore of
use to be able to control the number of calls to \disent{}. This is
done through the flag {\tt -ncall2}, introduced already above. 

We provide a small script to run on multiple cores on a single machine
in \masterlink{aux/runpar.sh}.

\subsection{Validating the code}
The code comes with a script to validate that the results come out as
expected. It can be run by entering \masterlink{validation} and
executing
\begin{lstlisting}
  ./validate_or_generate.sh validate
\end{lstlisting}
The script builds the code and executes a number of tests designed to
check the most important features of \disorder{}, including both the
NC and CC channels, the \fastjet{} and \lhapdf{} interfaces, the
automated PDF and scale uncertainty features, and P2B. The script
assumes default installation paths for \hoppet{}, \fastjet{}, and
\lhapdf{}. A user can however manually change this by inspecting the
script and adding the appropriate {\tt cmake} flags at the
beginning. The validation script runs in about 500 CPU seconds, or no
more than a few minutes on a modern laptop, utilising the {\tt
  parallel} program~\cite{tange_2021_5233953}.

\section{Benchmarks and results}
\label{sec:results}
\begin{table}[t] 
  \centering
  \phantom{x}\medskip
  \begin{center} $Q = 10\GEV,\quad x = 0.01$\end{center}
  {\renewcommand{\arraystretch}{1.2}
  \begin{tabular}{lcccc|cc}
    \toprule
    & $\frac{\sigma_\NC}{\dd x \dd Q^2}$ [pb/GeV$^2$] & $\delta$PDF & $\frac{\sigma_\CC}{\dd x \dd Q^2}$ [pb/GeV$^2$] & $\delta$PDF & $\tilde{\sigma}_\NC$  & $\tilde{\sigma}_\CC$ \\ 
    \midrule
    a\NNNLO & $1932^{+.705\%}_{-.486\%}$ & $0.972\%$ & $1.100^{+.591\%}_{-.423\%}$ & $1.103\%$ & $0.8174$ & $2.884$\\    
    \NNNLO  & $1886^{+.546\%}_{-.160\%}$ & $0.845\%$ & $1.080^{+.455\%}_{-.132\%}$ & $0.914\%$ & $0.7980$ & $2.829$\\    
    NNLO    & $1895^{+1.50\%}_{-1.18\%}$ & $0.840\%$ & $1.084^{+1.25\%}_{-.994\%}$ & $0.908\%$ & $0.8018$ & $2.840$\\    
    NLO     & $1952^{+3.66\%}_{-4.43\%}$ & $0.810\%$ & $1.111^{+3.02\%}_{-3.63\%}$ & $0.880\%$ & $0.8260$ & $2.913$\\    
    LO      & $2058^{+13.8\%}_{-17.0\%}$ & $0.843\%$ & $1.163^{+11.4\%}_{-14.1\%}$ & $0.885\%$ & $0.9708$ & $3.050$\\    
    \bottomrule
  \end{tabular}}
  \caption{The inclusive and reduced cross sections at various orders
    in both NC and CC DIS. The setup is given in
    eq.~\eqref{eq:setup}. The a\NNNLO{} row is obtained using the
    approximate \NNNLO{} PDF set MSHT20an3lo\_as118. Note that there
    is no Monte Carlo error on these numbers.}
\label{tab:cross-sections}
\end{table}
In this section we show a few results obtained by running the code in
both the inclusive and the P2B modes. The purpose is not to provide an
exhaustive phenomenological analysis, but rather show the capabilities
of \disorder{} and to provide a few select results that can be used to
either validate the code when running it, or to validate other
programs in the future. All analyses used in this section can be found
in the \masterlink{analysis} directory, and the raw results can be found in
the \masterlink{paper\_runs} folder.

We use {\tt fastjet~v3.4.1} for jet-clustering, and unless otherwise
stated we collide electrons and protons using the
MSHT20nnlo\_as118~\cite{Bailey:2020ooq} PDF set using {\tt
  LHAPDF~v6.5.4} and the following input parameters in all
calculations
\begin{align}
  \label{eq:setup}
  M_W = 80.398\GEV, \qquad M_Z = 91.1876\GEV, &\qquad E_h = 920\GEV, \qquad E_l = 27.6\GEV,  \\ \notag
  \alpha = 1/137, \qquad \as(M_Z) &= 0.118, \qquad n_f = 5\,.
\end{align}

\subsection{Inclusive results}
The perhaps most fundamental question one can ask in DIS is ``what is
the total cross section for a given value of $x$ and $Q$?''. The
answer to that question is given in table~\ref{tab:cross-sections} for
both the NC and CC channels at $Q=10\GEV$ and $x=0.01$, at all
available perturbative orders.  The row a\NNNLO{} is obtained with the
MSHT20an3lo\_as118~\cite{McGowan:2022nag} PDF set which includes
approximate \NNNLO{} theoretical input. It is interesting to note that
the inclusion of the approximate terms has an impact that is
parametrically of the same order as the NNLO corrections. The PDF
uncertainties stay more or less constant across all orders, which is
not unexpected given that we use the same PDF at each order. To give a
sense of the speed of the code, the combined results presented in the
table, which corresponds to 325 different PDF members and hence
structure functions, took a total of 3.5 minutes to obtain on a laptop
equipped with an Intel i9-10885H CPU.
\begin{figure}[tb!]
  \centering\includegraphics[width=0.49\textwidth,page=1]{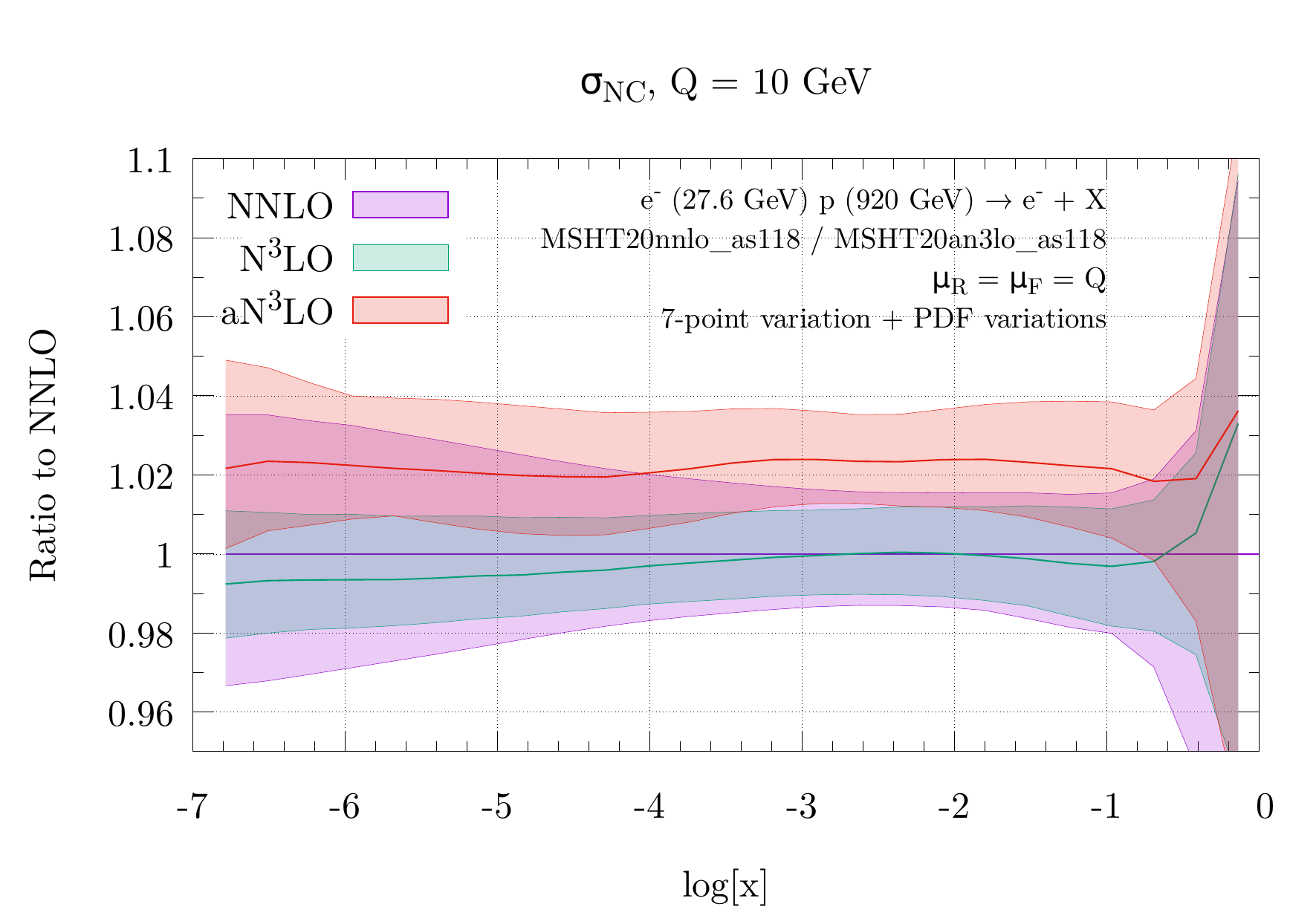}
  \centering\includegraphics[width=0.49\textwidth,page=3]{figures/sigma-ratios.pdf}
  \caption{The NC cross section for fixed $Q=10\GEV$ differential in
    $\log x$ (left) and for fixed $x=0.01$ and differential in $\log
    Q$ (right). We show NNLO (purple), \NNNLO{} (green), and a\NNNLO{}
    (red). The uncertainty band is obtained as the linear combination
    of scale and PDF uncertainties.}
  \label{fig:sigma}
\end{figure}

In figure~\ref{fig:sigma} we show the NC cross section for fixed
$Q=10\GEV$ (left) or $x=0.01$ (right), now only at NNLO, \NNNLO{} and
a\NNNLO{}. The uncertainty bands are obtained as a linear combination
of the scale uncertainty and the PDF uncertainty, and is typically
dominated by the latter, in particular at \NNNLO{}. We observe that
although the central prediction at a\NNNLO{} is in tension with the
\NNNLO{} curve, it is contained within the NNLO scale uncertainty
band, except at intermediate $x$ values. It will be interesting to
see, if the inclusion of the exact \NNNLO{} splitting functions, when
they become available, will ameliorate this tension.

\subsection{Exclusive laboratory frame results}
To demonstrate the results of the code in P2B mode, we perform a jet
analysis in the laboratory frame, based on Ref.~\cite{Borsa:2022cap},
which probes the kinematic range accessible by the upcoming EIC. For
this analysis we set $E_l=18\GEV$, $E_h=275\GEV$, and restrict the DIS
kinematics by
\begin{align}
  25\GEV^2 < Q^2 < 1000 \GEV^2, \quad 0.04 < y < 0.95\,.
\end{align}
We reconstruct jets in the laboratory frame using the anti-$k_T$
algorithm~\cite{Cacciari:2008gp} with $R=0.8$. Jets are those that
satisfy the following transverse momentum and pseudo-rapidity requirements
\begin{align}
  p_{t,j} > 5 \GEV, \quad |\eta_j| < 3\,.
\end{align}
\begin{figure}[tb!]
  \centering\includegraphics[width=0.49\textwidth,page=1]{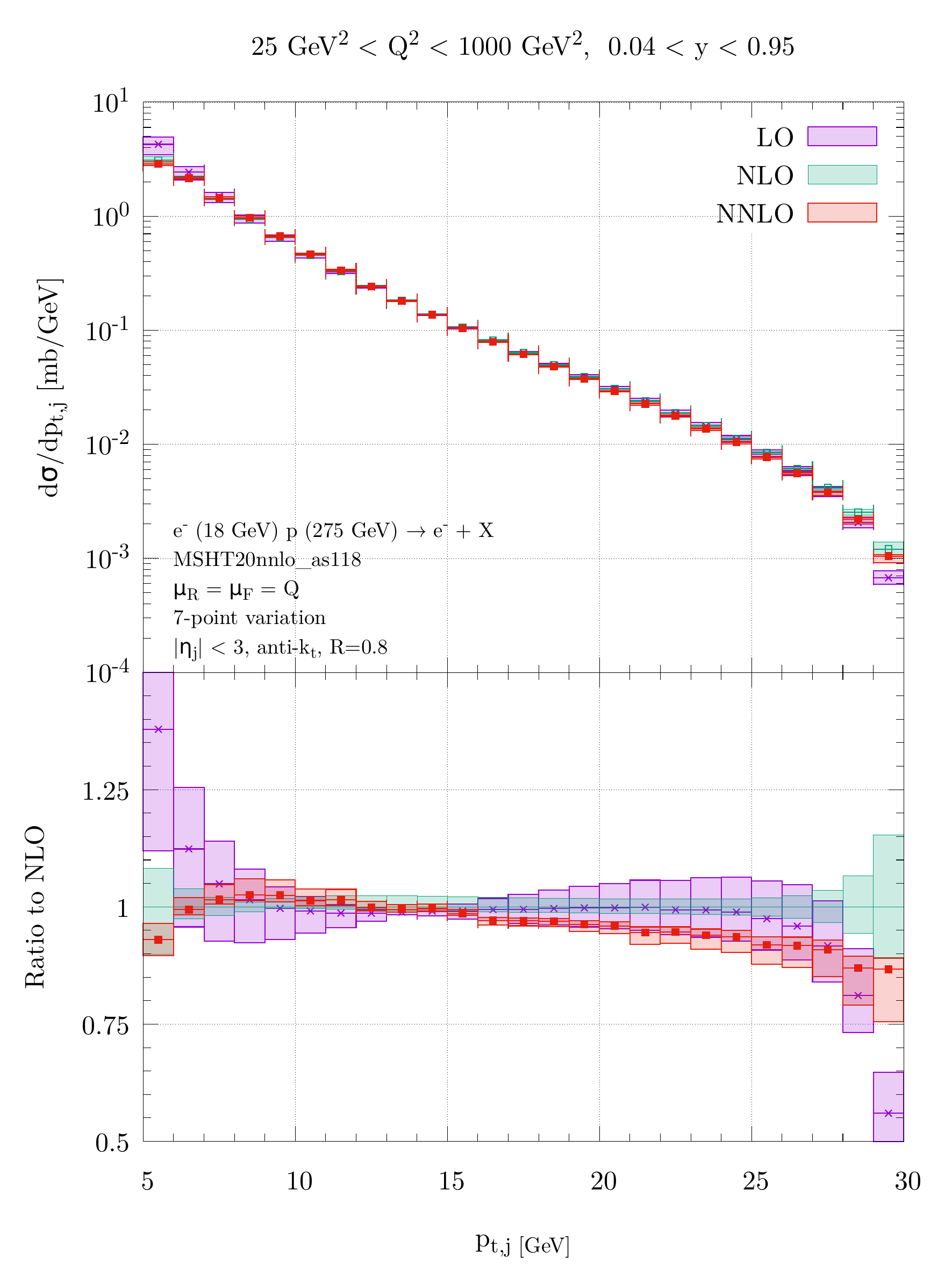}
  \centering\includegraphics[width=0.49\textwidth,page=2]{figures/lab-frame.pdf}
  \caption{The NC cross section for fixed $Q=10\GEV$ differential in
    $\log x$ (left) and for fixed $x=0.01$ and differential in $\log
    Q$ (right). We show NNLO (purple), \NNNLO{} (green), and a\NNNLO{}
    (red). The uncertainty band is obtained as the linear combination
    of scale and PDF uncertainties.}
  \label{fig:labframe}
\end{figure}
In Fig.~\ref{fig:labframe} we show the inclusive jet transverse
momentum and rapidity at LO, NLO, and NNLO, in photon-induced NC
DIS. For the transverse momentum the higher order corrections are
moderate except for small values and close to the LO threshold of
$p_{t,j}=Q$. The rapidity distribution on the other hand receives
larger corrections. This is in part due to hard jets with $p_{t,j}>Q$
that can only arise at NLO and beyond. Here the scale variation bands
do not cover the actual corrections, and it is obvious that one needs
to go beyond NLO for reliable predictions. Of course one needs the
full \NNNLO{} to assess the reliability of the NNLO result. However,
from similar results in Ref.~\cite{Currie:2018fgr} it can be expected
that the actual \NNNLO{} results will be mostly contained within the
scale variation band at NNLO.

\section{Conclusion}
\label{sec:conclusion}
In this paper we have presented version 1.0.0 of a Fortran code,
\disorder{}, capable of computing massless DIS cross sections fully
differentially at NNLO and inclusively at \NNNLO{}. The code combines
the NNLO structure functions from \hoppet{} and the NLO dijet
calculation of \disent{}, using the P2B method, thereby obtaining
fully differential single-jet DIS predictions in the laboratory
frame. In the inclusive mode the program includes all NC and CC
contributions and can carry out both scale and PDF variations on the
fly. Due to the underlying tabulation of the structure functions in
\hoppet{}, the code is extremely fast, and a user can obtain cross
section results at \NNNLO{} in a matter of seconds.

In the fully differential P2B mode, only photon-mediated DIS can be
computed, and no PDF variations are currently implemented. It should
in principle be possible to extend \disent{} to full NC and CC, and
this is something that is planned for a future release of the
code. The extension to CC would in fact be very beneficial, as the
Breit-frame cannot be reliably determined due to the undetected
neutrino. Similarly the PDF variations could also be implemented,
however this would require some serious restructuring of the code for
it to be efficient. Extension of the code to handle incoming neutrino
beams in the P2B mode, of relevance to the FASER~\cite{Feng:2017uoz}
physics program, is also planned.

The upgrade of the code to handle \NNNLO{} fully differentially is
however not currently planned. This is mainly due to the fact that
this would require a complete replacement of \disent{} with a proper
NNLO dijet code, entailing a significant amount of
work. Such a code would also be very slow compared to what is in place
now, and it would defeat part of the attraction of \disorder{} which
is its speed. At LO and NLO a user can obtain differential results on
a laptop running for a just a few minutes, and even reasonable NNLO
results can be obtained on a multi-core machine running for an
hour. \NNNLO{} results would inevitably require large resources on a
High Performance Cluster.

Given the code's dependency on \hoppet{}, any developments therein
will almost automatically propagate to \disorder{}. In particular it
is expected that the exact $\mathcal{O}(\as^4)$ splitting functions,
as opposed to the approximations which are currently employed, will
become available in \hoppet{} once they have been fully
determined. This is currently the only missing piece needed to
formally claim full \NNNLO{} accuracy for the structure
functions. Similarly, were the massive DIS coefficient
functions~\cite{Gottschalk:1980rv,Laenen:1992zk,Laenen:1992xs,Gluck:1997sj,Blumlein:2011zu,Behring:2015roa,Berger:2016inr,Gao:2017kkx}
to be implemented in \hoppet{} the structure functions in \disorder{}
could immediately be modified to accommodate this.

The code is intended to be user-friendly and to this effect comes with
an interface to \fastjet{} and \lhapdf{} and is run through a
command line interface. 

\section*{Acknowledgements}
The author is grateful for many discussions with Andrea Banfi and
Gavin Salam regarding \disent{}, and for encouragement to publish
\disorder{}. The author also thanks Valerio Bertone for discussions
about the DIS structure functions and Alex Huss for valuable cross-checks
against results obtained with the \nnlojet{} code. The idea for the
code originated while the author was working on
Ref.~\cite{Banfi:2023mhz}.

%
%

\bibliography{bibliography.bib}

\nolinenumbers

\end{document}